# Capacitated Vehicle Routing Problem Using Conventional and Approximation Method

Apurv Choudhari
Department of Computer Engineering
Vishwakarma Institute of Technology
Pune, India
apurv.choudhari16@vit.edu

Ameya Ekbote
Department of Computer Engineering
Vishwakarma Institute of Technology
Pune, India
ameya.ekbote16@vit.edu

Prerona Chaudhuri
Department of Computer Engineering
Vishwakarma Institute of Technology
Pune, India
prerona.chaudhuri16@vit.edu

*Abstract*—This paper attempts to solve the famous Vehicle Routing Problem by considering multiple constraints including capacitated vehicles, single depot, and distance using two approaches namely, cluster first and route the second algorithm and using integer linear programming. A set of nodes are provided as input to the system and a feasible route is generated as output, giving clusters of nodes and the route to be traveled within the cluster. For clustering the nodes, we have adopted the DBSCAN algorithm, and the routing is done using the approximation algorithm, Christofide's algorithm. The solution generated can be employed for solving real-life situations, like delivery systems consisting of various demand nodes.

## I. INTRODUCTION

Vehicle routing problem (VRP), defined by Dantzig and Ramsar for the first time in literature, is a problem to determine an optimal vehicle route to serve from one or more depots to ''n'' number of customers. Today, distribution systems have become quite complicated and increased the transportation cost of companies. When a vehicle routing plan is made effectively, it can provide significant savings in the cost. The objective of VRP is to minimize traveling distance and/or time among customers. A vehicle should travel an optimal path visiting all nodes with satisfying the constraints. The difficulty of the problem varies with the addition of different constraints to the VRP.

Many different variants of vehicle routing problem have been extensively studied. All these variants consider different constraints while solving the optimization problem. Some of these constraints are vehicle capacity, the number of vehicles, maximum distance traveled by the vehicle, time window for each customer node, etc. Here we have considered two constraints 1) capacity of the vehicle and 2) maximum distance traveled. Many different approaches can be applied to solve the problem. Some of them are using heuristics/metaheuristics algorithms, genetic algorithms, evolutionary algorithms, exact algorithms, approximation algorithms, etc. Finding the optimal solution for this problem takes exponential time. Most of the variants of VRP are NP-Hard. So, we have used approximation algorithm which solves the problem in polynomial time. The approach used here is cluster first route second.

## II. LITERATURE REVIEW

Vehicle routing can be defined as the distribution of products to specific customers from several warehouses and the collection of products from different customers in a logistic system. The VRP is a generalized version of the Travelling Salesman Problem (TSP), and
also, the VRPs proved to be non-deterministic polynomial-time (NP)-hard problem [4]. It was brought for the first time into literature by Dantzig and Ramsar in 1959. In their study, the authors focused on the distribution of gasses to petrol stations by establishing the first mathematical programming model. Later, a saving algorithm that is a heuristic algorithm for a routing problem was proposed by Clarke and Wright [5]. An inclusive literature review related to the VRP was made by Toth and Vigo [6]. There is extensive research related to VRP based on different constraints and types. However, in this section, only some of the studies related to VRP in the literature are summarized.

### 1. CLARKE-WRIGHT SAVINGS ALGORITHM

Clarke & Wright algorithm an algorithm used to provide solutions to vehicle routing problems was introduced in the year 1964, this algorithm mainly uses the concept of Savings to determine the solution. A distance matrix is prepared which provides us the details of the distance between various nodes present in the city and also the distance between the nodes and the warehouse also. The cost can be used in place of distance if the transportation cost between every pair of the node is available to us.

The distance $D_{ij}$ on a grid between a point $i$ with a coordinate $(x_i, y_i)$ and a point $j$ with a coordinate $(x_j, y_j)$ is evaluated as:
$$D_{ij} = \sqrt{(x_i - x_j)^2 + (y_i - y_j)^2}$$
In this algorithm, the individual vehicle serves the available 'n' nodes. Then it calculates the savings for joining the cycles using the edge $[i, j]$, using the formula:
$$S_{ij} = D_{0i} + D_{0j} - D_{ij}$$
As there are n nodes total $C_2^n$ number of savings will be generated using the above equation. Sort the savings in descending order. Now keep assigning the next highest saving for a vehicle by satisfying capacity constraints. If any node does not satisfy capacity constraints, then discard that node and consider the next node which satisfies the capacity constraints. In this way assign all the nodes to the vehicles without satisfying capacity constraints.

## 2. CLUSTERING METHODS

### 2.1 Capacitated K-means clustering algorithm

K-means, developed by MacQueen in 1967, is one of the clustering algorithms [58]. The formal logic of K-means is to divide data into K clusters. The aim of the clustering analysis is based on the principle of maximizing the similarity of groups within themselves and minimizing between-group similarity. Cluster similarity is measured by the average distance between cluster center points and the dataset point. The algorithm needs a fixed number of clusters. Therefore, the algorithm is called K-means. K represents the number of clusters and is previously known, as a constant positive integer number. K-means algorithm gives acceptable results with all types of data even though some clustering algorithms achieve better results in some series. A disadvantage of the algorithm is determining the K values. Therefore, the optimal K number must be found by a trial and error to achieve a successful clustering.

### 2.2 Capacitated K-medoids clustering algorithm

K-medoids algorithm was developed by Kaufman and Rousseeuw to reduce the sensitivity of K-means in data with noise and exception data. The element at most center points in the algorithm is accepted as a new cluster center. Thus, the exception data are not allowed to scroll cluster center towards edges. There are many different types of K-medoids algorithm. Partitioning Around Medoids (PAM) is the first K-medoids algorithm asserted. K number is accepted as the cluster's center for the PAM algorithm. When a new element (point) is added to the cluster, if a point contributes to the cluster, this point must be accepted as a new center of the previous old center. When a point's contribution to the cluster's improvement is confirmed, this point may be a new center, and an old center is accepted as a simple element of the cluster. Despite the fact that PAM provides very good results in small databases, it shows underperformance in complex databases.

### III. SOLUTION APPROACH

We have used two approaches to solve the same problem. The first approach is cluster first route second approach to solve the problem. We will generate clusters of customer nodes based on the given constraints. Then depot node will be considered in each cluster explicitly. The next step is to find the optimal path in each cluster where the vehicle starts traveling from the depot and returns to the depot again satisfying the needs of all the customer nodes. For clustering, the DBSCAN algorithm is used, and to find the optimal path Christofide's algorithm, which is an approximation algorithm used to solve TSP in polynomial time with approximation factor 3/2, is used.

The second approach is by formulating the actual problem and solving it using Integer Linear Programming (ILP).

### A. PROBLEM FORMULATION:

*Integer Linear Programming*

The goal is to determine a route schedule that minimizes the traveled distance and the number of vehicles:

n: number of clients
N: the set of clients
V: the set of vertices (considering depot i.e., node 0 along with N)
A: set of arcs, $\{(i,j) \in V2, i \neq j\}$
$C_{ij}$: cost of travel over arc $(i,j) \in A$
Q: maximum capacity of the vehicle
$q_i$: quantity to be delivered to node $i \in N$

to achieve: Minimize $\sum_{ij \in A} C_{ij} X_{ij}$
such that:
$$\sum_{j \in V, i \neq j} X_{ij} = 1 \quad i \in N$$
$$\sum_{i \in V, i \neq j} X_{ij} = 1 \quad j \in N$$
If $X_{ij} = 1$ then
$$u_i + q_j = u_j, i,j \in A: j \neq 0, i \neq 0$$
$$q_i \leq u_i \leq Q \quad i \in N$$
$$x_{ij} \in \{0,1\} \quad ij \in A$$

To solve the above constraints, an optimization studio by IBM, named CPLEX has been used. ILP is NP-hard since in the worst case it can consider all the possible integer vectors ($2^n$), hence it will always be providing an optimized solution over the method mentioned below – cluster first and route second.

### B. DBSCAN

Density Based Spatial Clustering of Applications with Noise (DBSCAN) algorithm, which is one of the density-based clustering methods, was developed by Martin Ester, Hans-Peter Kriegel, Jörg Sander, and Xiaowei Xu. In DBSCAN center-based approach, the density for a point in the database is predicted by considering the number of points located within an exact distance (the longest neighborhood radius, radius of Eps) from this point. This point number contains the point itself. Clusters are formed by applying capacity and distance constraints (for finding out nearest neighbors) for which three parameters have been considered for implementing the DBSCAN algorithm: Eps, MinWt and MaxWt. Eps is the longest neighborhood radius. MinWt represents the minimum capacity (or weight) of elements located in a neighborhood region of radius Eps, which will be carried by a vehicle. MaxWt is the maximum capacity (or weight) that can be carried by the vehicle. The aim of clustering is to find out the optimum number of vehicles used along with the nodes in various clusters. Here all the vehicles are considered of the same capacity. The steps of the DBSCAN algorithm are explained below. The DBSCAN is a very useful clustering method, especially for large databases and databases that contain noisy data. It is often used for clusters with different volumes and shapes.

In DBSCAN initially, Eps radius neighborhood regions of each element in the database are searched. In this region, if the capacity (or weight) of neighbor points is bigger than MinWt, the element and its neighbors are accepted as a new cluster. Then again, each node in the cluster is considered and its neighbors are found out until the maximum weight (or capacity) is not exceeded. The process is terminated if no new node is added in the cluster.

## C. Christofide's Algorithm

The clustering algorithm will give the sets of nodes satisfying all the constraints considered here. Next step is to find optimal path for vehicle in each cluster where vehicle starts from and ends at depot. One of the best methods to solve this is solving Travelling Salesman Problem for each cluster. In this approach, we have used Christofide's algorithm which is an algorithm for finding approximate solution of Travelling Salesman Problem, which works on metric space. It ensures the solution is within a factor of 3/2 of the optimal solution.

The algorithm states that, let $G = (V, w)$ be an instance of traveling salesman problem. Here G is a complete graph with set V of vertices and w is the non-negative real weight for each edge of G. It finds a near-optimal solution using following steps.

1. Create a minimum spanning tree T from the given graph G.
2. Let set O of odd degree vertices from T. There will be even number of odd degree vertices.
3. Find minimum weight perfect matching M in induced subgraph given by vertices in O.
4. Combine edges of M and T to form a multigraph H with all even degree vertices.
5. Form an Eulerian circuit from H.

We will solve TSP for each cluster generated using DBSCAN. It will give an optimal path in each cluster, which is for each vehicle, where trip starts from and ends at depot.

The cost of the solution produced by the algorithm is within 3/2 of the optimum. To prove this, let C be the optimal traveling salesman tour. Removing an edge from C produces a spanning tree, which must have weight at least that of the minimum spanning tree, implying that w(T) ≤ w(C). Next, number the vertices of O in cyclic order around C, and partition C into two sets of paths: the ones in which the first path vertex in cyclic order has an odd number and the ones in which the first path vertex has an even number. Each set of paths corresponds to a perfect matching of O that matches the two endpoints of each path, and the weight of this matching is at most equal to the weight of the paths. Since these two sets of paths partition the edges of C, one of the two sets has at most half of the weight of C, and thanks to the triangle inequality its corresponding matching has weight that is also at most half the weight of C. The minimum-weight perfect matching can have no larger weight, so w(M) ≤ w(C)/2. Adding the weights of T and M gives the weight of the Euler tour, at most 3w(C)/2. Thanks to the triangle inequality, shortcutting does not increase the weight, so the weight of the output is also at most 3w(C)/2.

## RESULTS

The results of both approaches have been demonstrated below. It can be observed that integer linear programming generates a smaller number of clusters than the approximation method of cluster first and route second.

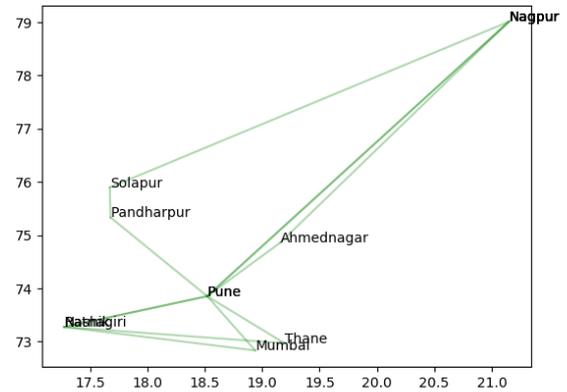

Using cluster first and route second

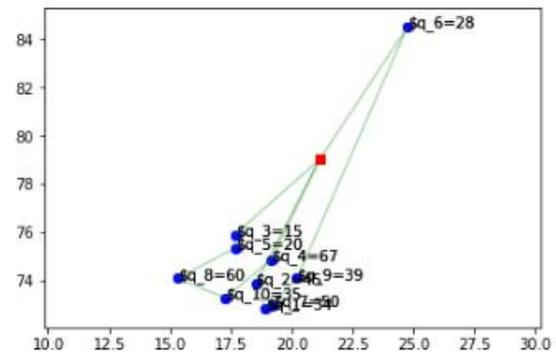

Using integer linear programming

## FUTURE SCOPE

In further development, we will implement a Recursive DBSCAN clustering algorithm. In order to satisfy all demand nodes, consideration of noise nodes is important. This algorithm will help us to effectively solve the problem of noise nodes. Also, DBSCAN may find it difficult to cluster high-density regions which can be effectively done by recursive DBSCAN.

The difficulty of the problem varies with the addition of different constraints to the VRP. Along with distance, the capacity of vehicle as well as demand of customer nodes we will also consider some real-life constraints like hard time window and soft time window. When the constraints of the earliest time and the latest time to start a service are added to the VRP, the problem becomes a vehicle routing problem with time windows (VRPTW). The VRPTW is a type of VRP where the customers can be served in a time window $[a_i, b_i]$ ; $a_i$ and $b_i$ represents the earliest and latest time to start the service. The objective of VRPTW is to design an optimal route using vehicle capacity, service time, and time windows. Vehicles that arrive earlier than its lower bound have to wait and no additional cost is incurred but vehicles that arrive later than its upper bound are prohibited from serving in the VRPHTW.

ACKNOWLEDGMENT

We have extremely benefitted from fruitful discussions with, Prof Dr. Pushkar Joglekar and Prof Karthick Subramaniam. We also want to thank the Computer Department of VIT, for encouraging us for undertaking this real-world problem.